\documentclass[fleqn,twoside]{article}
\usepackage{espcrc2}

\usepackage{graphicx}
\usepackage{epsfig}

\newcommand{\psp}{\psi(2S)}

\newcommand{\pppsp}{\pi^+\pi^- \psp}

\newcommand{\ee}  {\ensuremath{ e^+ e^- }}
\newcommand{\ecm} {\ensuremath{ E_{\mathrm{c.m.}} }}
\newcommand{\sqs} {\ensuremath{ \sqrt{s} }}

\newcommand{\RM}  {\ensuremath{ M_{\mathrm{rec}} }}

\newcommand{\RMS} {\ensuremath{ M^2_{\mathrm{rec}} }}

\newcommand{\gisr}{\ensuremath{\gamma_{\mathrm{isr}}}}
\newcommand{\gevc}{\ensuremath{\, {\mathrm{GeV}/c^2} }}
\newcommand{\mevc}{\ensuremath{\, {\mathrm{MeV}/c^2} }}
\newcommand{\gev} {\ensuremath{\, {\mathrm{GeV}} }}

\newcommand{\eeddst}     {\ensuremath{e^+e^- \to D^{(*)\pm}{D}{}^{*\mp} }}
\newcommand{\ddstch}     {\ensuremath{            D^{(*)+} D^{*-}      }}
\newcommand{\eeddstch}   {\ensuremath{ e^+e^- \to D^{(*)+} D^{*-}      }}
\newcommand{\eeddstchg}  {\ensuremath{ e^+e^- \to D^{(*)+} D^{*-} \gisr}}
\newcommand{\eedpdstm}   {\ensuremath{ e^+e^- \to D^+    D^{*-} }}
\newcommand{\eedstpdstm} {\ensuremath{ e^+e^- \to D^{*+} D^{*-} }}

\newcommand{\dd}    {\ensuremath{D \overline D }}

\newcommand{\eedd}  {\ensuremath{ e^+e^- \to D \overline D            }}
\newcommand{\eeddg} {\ensuremath{ e^+e^- \to D \overline D \gamma_{ISR}}}
\newcommand{\eeddb} {\ensuremath{ e^+e^- \to D^0 \overline D{}^0 }}
\newcommand{\eedpdm}{\ensuremath{ e^+e^- \to D^+ D^- }}

\newcommand{\psiddt} {\ensuremath{\psi(4415)\to D \overline D{}^{*}_2(2460) }}
\newcommand{\eeddp}  {\ensuremath{ e^+e^- \to D^0 D^- \pi^+ }}
\newcommand{\eeddpg} {\ensuremath{ e^+e^- \to D^0 D^- \pi^+ \gamma_{ISR} }}
\newcommand{\ddp}    {\ensuremath{ D^0 D^- \pi^+ }}

\newcommand{\ps}    {\ensuremath{ \psi(4415) }}

\newcommand{\dt}    {\ensuremath{ \overline D{}^{*}_2(2460)   }}

\newcommand{\ddt}   {\ensuremath{ D  \overline D{}^{*}_2(2460)   }}

\newcommand{\mddp}  {\ensuremath{ M_{D^0 D^-\pi^+} }}

\newcommand{\mdn}   {\ensuremath{ M_{D^0}   }}
\newcommand{\mdm}   {\ensuremath{ M_{D^-}   }}

\title{Charm cross-section and charmonium(like) states in \ee\ annihilation with Belle}

\author{T.~Uglov
\address[MCSD]{Institute for Theoretical and
    Experimental Physics,
\\ Bolshaya Cheremushkinskaya, 25, 117218,
    Moscow, Russia}}

\begin{document}

\begin{abstract}
We report BELLE measurements of the exclusive cross sections for the
processes \eeddst, \eedd, \eeddp, the first observation of
\psiddt\ decay and new state, $Y(4660)$, using ISR. In addition,
another cluster of events at around 4.05~GeV/$c^2$ is reported.
\vspace{1pc}
\end{abstract}

\maketitle

\section{Introduction}

Exclusive \ee\ hadronic cross sections to final states with charm
meson pairs are of special interest since they provide information
on the spectrum of $J^{PC}=1^{--}$ charmonium states above the
open-charm threshold. Parameters of these states obtained from fits to
the inclusive cross section~\cite{bes:fit} are poorly understood
theoretically~\cite{barnes}.  

Initial-state radiation (ISR) is proved to be a powerful tool for
measurement of the \ee\ exclusive hadronic cross sections at
\sqs\ smaller than the initial \ee\ center-of-mass energy (\ecm) at
$B$-factories.  ISR allows to obtain cross sections in a broad energy
range while the high luminosity of the $B$-factories compensates for
the suppression associated with the emission of a hard photon.

Here we report the first observation of the new charmonium-like state,
$Y(4660)$~\cite{y4660}, clustering structure near
4.05~GeV/$c^2$~\cite{y4008}, the first measurement of the exclusive
cross sections for the processes \eeddst\ ~\cite{belle:ddst},
\eedd\ ~\cite{belle:dd}, \eeddp\ and the first observation of
\psiddt\ decay~\cite{belle:dd2}.  The data sample corresponds to a
large integrated luminosity collected with the Belle
detector~\cite{det} at the $\Upsilon(4S)$ resonance and nearby
continuum at the KEKB asymmetric-energy \ee\ collider~\cite{kekb}.

\section{Recoil mass technique}

There are two ways of  ISR event reconstruction: partial or full.

In the full reconstruction method we select \eeddg\ signal
events by reconstructing both the $D$ and $\overline D$
mesons. In general, the \gisr\ is not required to be detected; its
presence in the event is inferred from a peak at zero in the spectrum
of the recoil mass against the \dd\ system. The square of the recoil
mass is defined as:
\begin{eqnarray}
\RMS(\dd)=(\ecm - E_{\dd})^2 - p^2_{\dd} , \nonumber
\end{eqnarray}
where $E_{\dd}$ and $p_{\dd}$ are the $\mathrm{c.m.}$ energy and
momentum of the \dd\ combination, respectively.

To select \eeddstchg\ signal events we use the partial reconstruction
method that achieves high efficiency by requiring full reconstruction
of only one of the $D^{(*)+}$ mesons, the \gisr, and the slow
$\pi_{\mathrm{slow}}^-$ from the other $D^{*-}$~\cite{foot1}. In this
case the spectrum of masses  recoiling against the $D^{(*)+} \gisr$
system
\begin{eqnarray}
 {\small\RM(D^{(*)+}\gisr)\!=\!\sqrt{(\ecm\!-\!E{}_{D^{(*)+} \gamma}){}^2\!-\!
p{}^2_{D^{(*)+} \gamma}}}\nonumber
\end{eqnarray}
peaks at the $D^{*-}$ mass. Here $E_{D^{(*)+} \gamma}$ and $p_{D^{(*)+}
  \gamma}$ are the c.m. energy and momentum, respectively, of the
$D^{(*)+} \gisr$ combination. This peak is expected to be wide and
asymmetric due to the photon energy resolution function  and higher-order
corrections to ISR. To disentangle the
contributions from different final states and to suppress combinatorial
backgrounds, we use the slow pion from the unreconstructed
$D^{*-}$. The difference between the mass recoiling against $D^{(*)+}
\gisr$ and $D^{(*)+} \pi_{\mathrm{slow}}^- \gisr$ (recoil mass
difference):
\begin{eqnarray}
{\small \Delta\RM \!= \!\RM(D^{(*)+}\gisr)\!-\!\RM(D^{(*)+}
\pi_{\mathrm{slow}}^- \gisr)\, ,}\nonumber
\end{eqnarray}
has a narrow distribution around the nominal $m_{D^{*-}}-
m_{\overline{D}{}^0}$ value, since the uncertainty in \gisr\ momentum
partially cancels out.

\section{Observation of the significant enhancement at 4.05\gev.}

We identify $\ee\to J/\psi\pi^+\pi^-\gisr$ process by peak in the
distribution on the  recoil mass  against the $J/\psi\pi^+\pi^-$ combination;
$J/\psi$ is reconstructed in $J/\psi\to\ee$ and $J/\psi\to\mu^+\mu^-$
modes.

\begin{figure}[h]
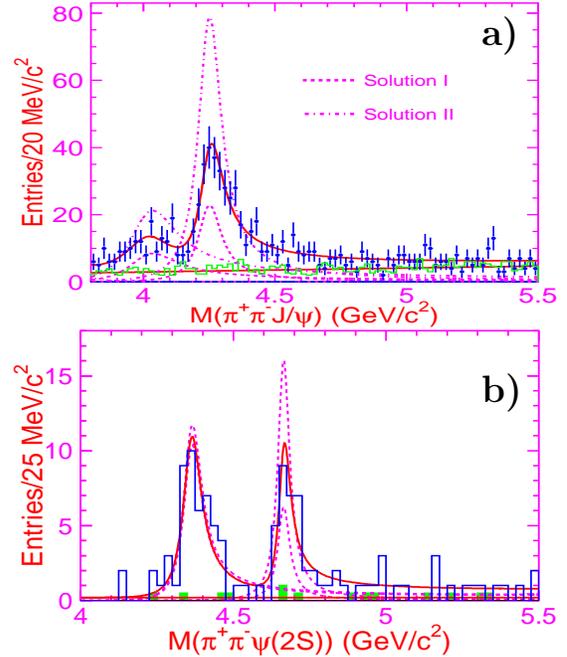

\begin{picture}(0,230)
\put(180,210){\Large \bf a)}
\put(180,75){\Large \bf b)}
\put(0,100){
\begin{minipage}[c]{0.45\textwidth}
\psfig{file=fig5_fit_plt.epsi,height=\textwidth,width=0.6\textwidth,angle=-90}


\psfig{file=fig2_solution12_plt.epsi,height=\textwidth,width=0.6\textwidth,angle=-90}
\end{minipage}
}
\end{picture}
\caption{Fit
to the a): $\pi^+\pi^-J/\psi$ b): $\pi^+\pi^-\psi(2S)$ mass spectrum with two coherent resonances. The
curves show the best fit and the contribution from each component.
The dashed curves are for solution I, and the dotted curves
for solution II.}
\label{coherent_two_res}
\end{figure}

Fig.~\ref{coherent_two_res} shows the $\pi^+\pi^-J/\psi$ invariant
mass distribution in the region of $3.8-5.5\gevc$.  There is a clear
enhancement at $4.25\gevc$ similar to that observed by the
BaBar Collaboration~\cite{babay4260}. In addition, there is a
clustering of events around $4.05\gevc$ that is significantly
above the background level.

An unbinned maximum likelihood fit is applied to the
$\pi^+\pi^-J/\psi$ mass spectrum in Fig.~\ref{coherent_two_res},~{\it a)}. Since
there are two clusters of events in the mass distribution, we fit it
with two coherent Breit-Wigner (BW) resonance functions assuming there
is no continuum production of $\ee\to \pi^+\pi^-J/\psi$. There are two
solutions with equally good fit quality.  The masses (($4008\pm
40^{+114}_{-28}$)\mevc\ and ($4247\pm 12^{+17}_{-32}$)\mevc\ for the
first and second states, respectively) and widths (($ 226\pm 44\pm
87$)\mevc\ and $ (108\pm 19\pm 10$)\mevc) of the resonances are the same
for both solutions; the partial widths to $\ee$ and the relative phase
between them are different. The interference is constructive for one
solution and destructive for the other.  The statistical significance
of the structure around 4.05~GeV/$c^2$ is estimated to be $7.4\sigma$
and is greater than $5\sigma$ in all of the fitting scenarios that are
considered.

\section{First observation of $Y(4660)$ state.}

Similar analysis is done for the $\pi^+\pi^-\psi(2S)\gisr$ final state.

Fig.~\ref{coherent_two_res},~{\it b)} shows the $\pppsp$ invariant mass for selected
$\psp$ events, together with background estimated from the scaled
$\psp$ mass sidebands.  Two distinct peaks are evident, one at
4.36~GeV/$c^2$ and another at 4.66~GeV/$c^2$.

An unbinned maximum likelihood fit that includes two coherent $P$-wave
Breit-Wigner (BW) functions and a constant, incoherent background is
applied to the $\pppsp$ mass spectrum in
Fig.~\ref{coherent_two_res},~{\it b)}. The fit results in two
solutions with equally good fit quality, masses (($4361\pm 9\pm
9$)\mevc\ for the first state, ($4361\pm 9\pm 9$)\mevc for the second state)
and widths (($74\pm 15\pm 10$)\mevc\ and ($48\pm 15\pm 3$)\mevc).  The
interference is constructive for one solution and destructive for the
other.  A statistical significance of $5.8\sigma$ is obtained for the
peak around $4.66\gevc$.

\section{Measurement  of the near-threshold \eeddst\ cross section}

\begin{figure}[htb]
\hspace*{-0.025\textwidth}
\includegraphics[width=0.48\textwidth,height=0.35\textwidth]{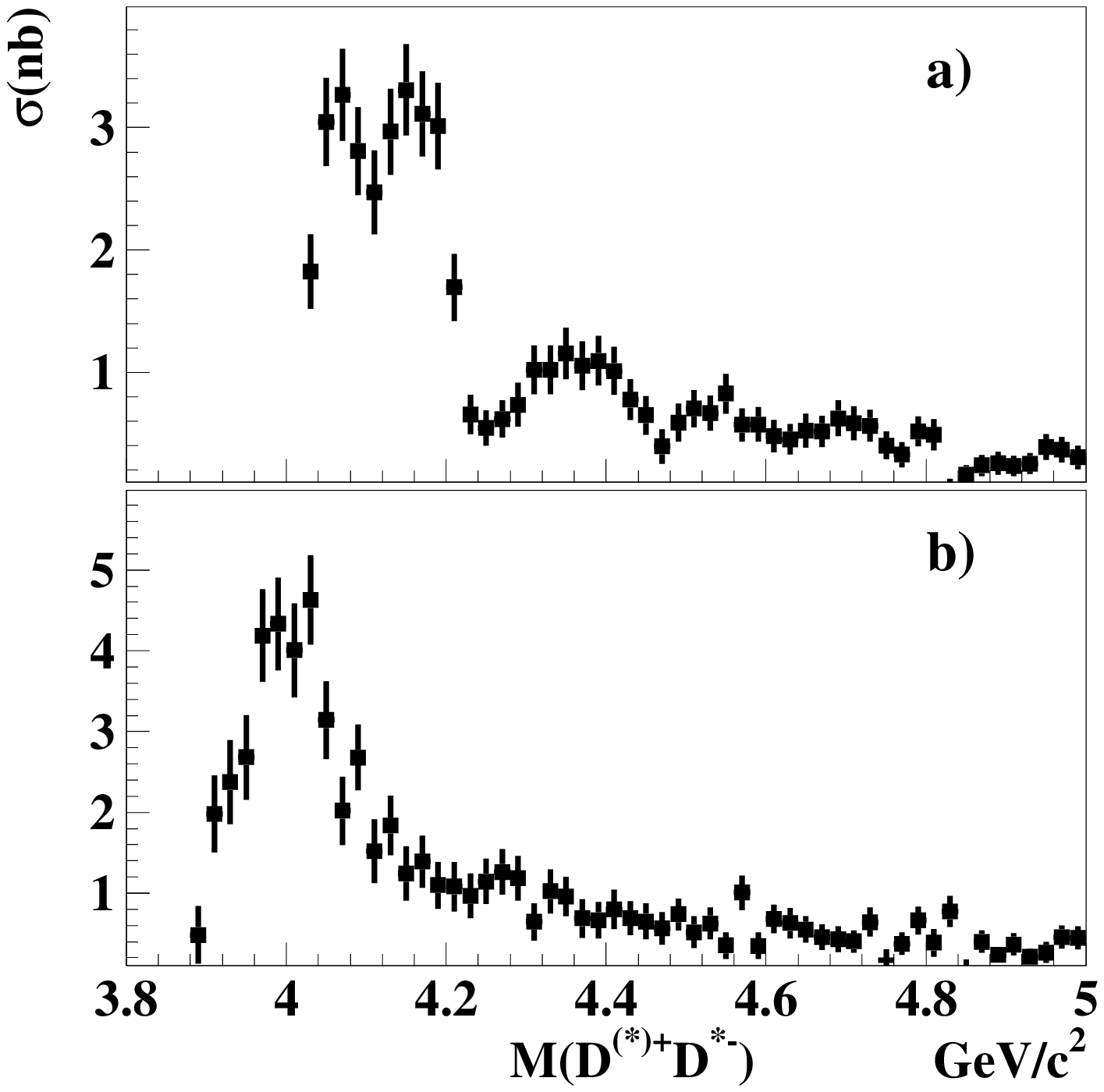}
\caption{The exclusive cross sections for a) \eedstpdstm\ and b) \eedpdstm.}
\label{Fig1}
\end{figure}

For the measurement of the exclusive cross section we determine the
\ddstch\ mass.  The \eeddstch\ cross sections are extracted from the
\ddstch\ mass distributions after background subtraction using the
relation described in~\cite{belle:ddst}.  The resulting exclusive
\eeddstch\ cross sections are shown in Fig.~\ref{Fig1}.
The shape of the \eedstpdstm\ cross section is complicated with
several local maxima and minima.  Aside from a prominent excess near
the $\psi(4040)$, the \eedpdstm\ cross section is relatively
featureless. The measured cross sections are compatible~\cite{foot2}
within errors with the $D^{(*)}\overline {D}{}^{*}$ exclusive cross
section in the energy region up to $4.260\gev$ measured by
CLEO-c~\cite{cleo:cs}.

\section{Measurement of the near-threshold \eedd\ cross section}

\begin{figure}[htb]
\hspace*{-0.025\textwidth}
\begin{picture}(0,230)
\put(0,100){
\begin{minipage}[c]{0.45\textwidth}
\psfig{file=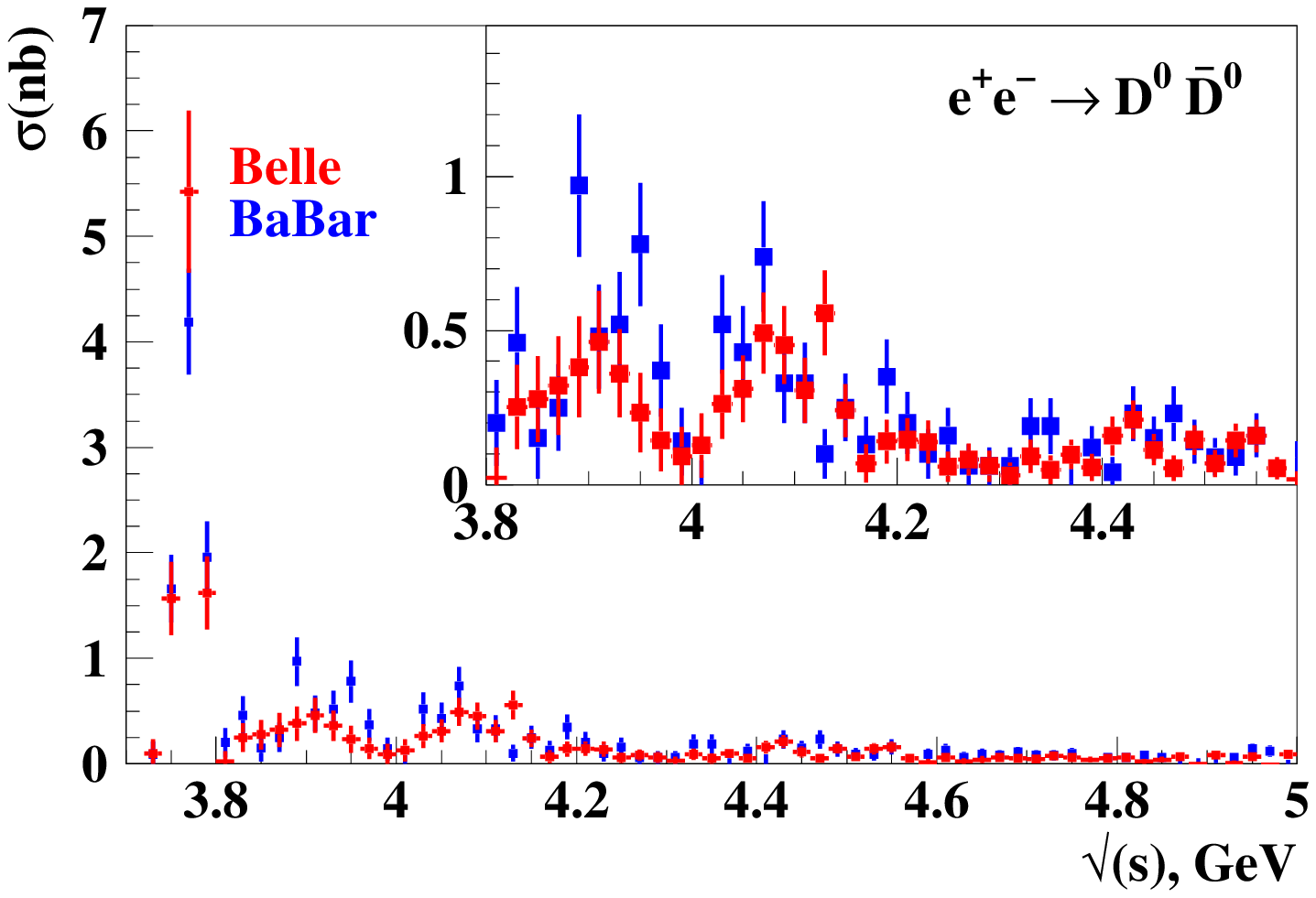,width=1.1\textwidth,height=0.5\textwidth}


\psfig{file=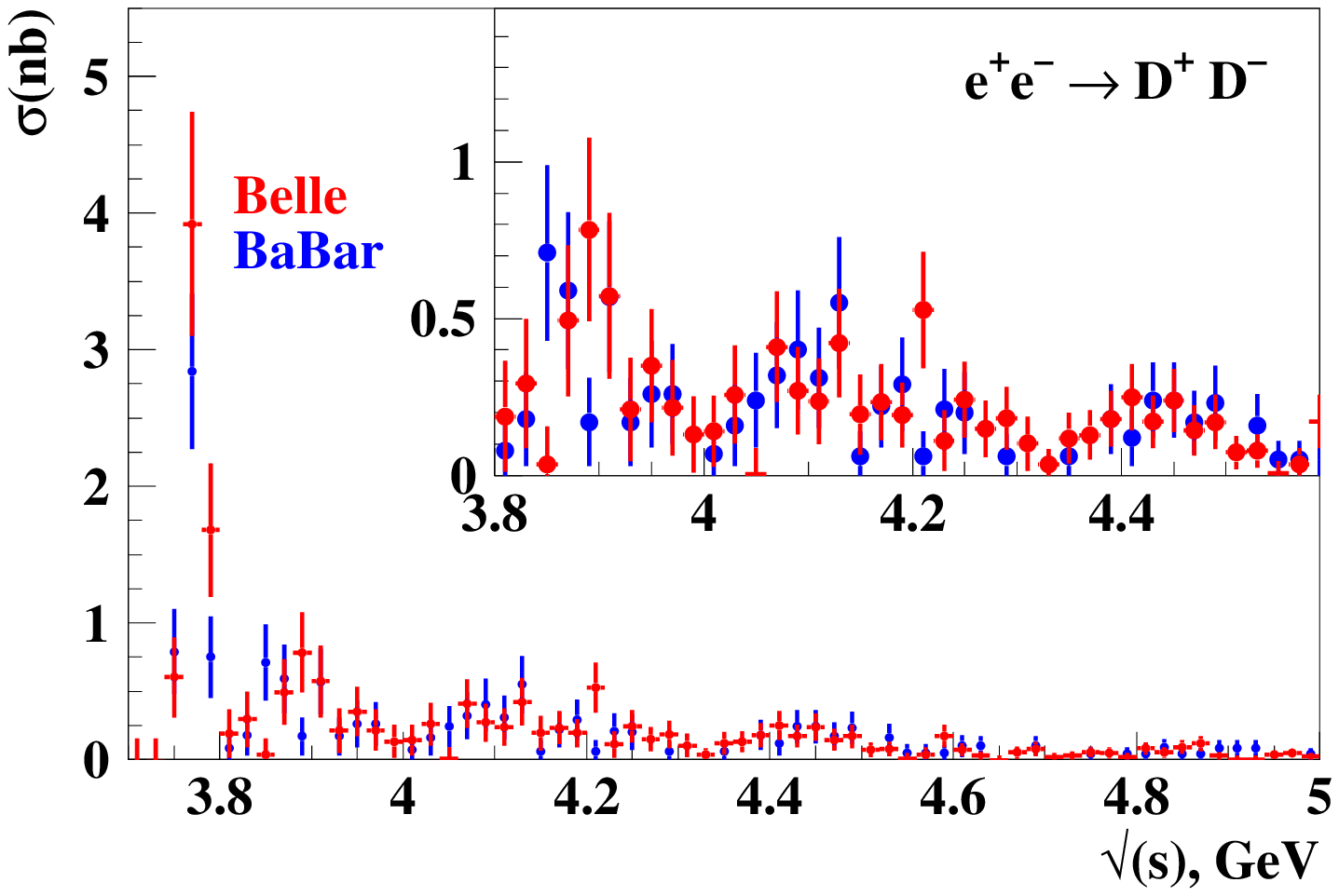,width=1.1\textwidth,height=0.5\textwidth}
\end{minipage}
}
\end{picture}
\caption{The exclusive cross sections for \eeddb\ and \eedpdm. The
  data are taken from the Durham database based on~\cite{belle:dd} and~\cite{babar:dd}.}
\label{Fig2}
\end{figure}

The \eeddb\ and \eedpdm\ exclusive cross sections, measured with full
event reconstruction method are shown in Fig.~\ref{Fig2}. Belle
results, shown with the red points, are compared to the BaBar data
(blue points).  The observed \eedd\ exclusive cross sections are
consistent with recent BaBar measurements~\cite{babar:dd} and are in
qualitative agreement with the coupled-channel model predictions of
Ref.~\cite{eichten}. This includes a peak at $3.9\gevc$ that is seen
both in Belle and BaBar mass spectra.

\section{Observation of \psiddt\ decay}
\begin{figure}[htb]
\begin{tabular}{cc}
\hspace*{-0.025\textwidth}
\includegraphics[width=0.4\textwidth]{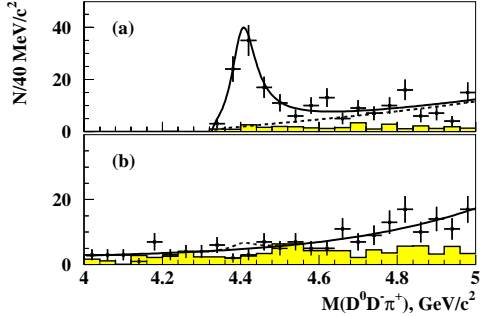}
\end{tabular}
\caption{(a) The \mddp\ spectrum for the \ddt\ signal region. The
  threshold function is shown by the dashed curve. (b) The
  \mddp\ spectrum outside the \ddt\ signal region.  The dashed curve
  shows the upper limit on the \ps\ yield at 90\% C.L.  Histograms
  show the normalized contributions from \mdn\ and \mdm\ sidebands.}
\label{Fig5}
\end{figure}

We use the  full reconstruction method described above to
select \eeddpg\ signal candidates. The \eeddp\ cross section extracted
from the background-subtracted \ddp\ mass distribution demonstrates a
prominent peak in a region of $\psi(4415)$ resonance.  To study the
resonant structure in \ps\ decays, we select \ddp\ combinations from a
$\pm 100\mevc$ mass window around the nominal \ps\ mass~\cite{pdg}. We
perform a separate study of $\ee\to\ddt$ and $\ee \to D(\overline D
\pi)_{\mathrm{non}\dt}$. The \mddp\ spectrum for the \ddt\ signal
interval is shown in Fig.~\ref{Fig5}(a). A clear peak corresponding to
\psiddt\ decay is evident near the \ddt\ threshold.  We perform a
likelihood fit to \mddp\ distribution with the \ddt\ signal
parametrized by an $s$-wave RBW function. The significance for the
signal is obtained to be $\sim10\sigma$. The obtained peak mass
$m_{\ps}=(4.411 \pm 0.007\mathrm{(stat.)})\gevc$ and total width
$\Gamma_{\mathrm{tot}}=(77 \pm 20\mathrm{(stat.)})\mevc$ are in good
agreement with the PDG~\cite{pdg} values, the recent BES
results~\cite{bes:fit} and predictions of Ref.~\cite{barnes}.

\section{Conclusions}

In summary, we presented the first observation of the new
charmonium-like state, $Y(4660)$, significant enhancement near
4.05\gevc, the first measurement of the exclusive cross sections for
the processes \eeddst, \eedd, \eeddp\ and the first observation of
\psiddt\ decay.  The obtained ISR results are in a good agreement with
the recent CLEO-c and BaBar measurements of the exclusive cross
sections.

\section{Acknowledgements}

The work is supported by President's Grant MK-4646.2009.2.


\begin{thebibliography}{9}
\bibitem{foot1} Charge-conjugate modes are included throughout this
  paper.
\bibitem{foot2} Since only charged final states are measured, our
  results should be scaled by a factor of two for this comparison.
\end{thebibliography}
\end{document}